\documentclass[aps,showpacs,nofootinbib,superscriptaddress]{revtex4}

\usepackage{graphicx}

\def\slashchar#1{\setbox0=\hbox{$#1$}
   \dimen0=\wd0 \setbox1=\hbox{/} \dimen1=\wd1
   \ifdim\dimen0>\dimen1 \rlap{\hbox to \dimen0{\hfil/\hfil}} #1
   \else  \rlap{\hbox to \dimen1{\hfil$#1$\hfil}} / \fi}
\def\jtstrut{\vrule height3ex depth0pt width0pt} 
\def\gev{\,\mathrm{GeV}}
\def\mev{\,\mathrm{MeV}}

\def\sth{s_\mathrm{th}}
\def\mi{\mathrm{i}}

\begin{document}

\title{Semileptonic $B\to \pi$ Decays from an Omn\`es Improved
Nonrelativistic Constituent Quark Model.}

\author{C. Albertus}\affiliation{Departamento de F\'{\i}sica Moderna,
  Universidad de Granada, E-18071 Granada, Spain.}
\author{J.M. Flynn}\affiliation{School of Physics \& Astronomy,
  University of Southampton, Southampton SO17 1BJ, UK}
\author{E. Hern\'andez} \affiliation{Grupo de F\'\i sica Nuclear,
  Facultad de Ciencias, E-37008 Salamanca, Spain.} 
\author{J. Nieves}\affiliation{Departamento de F\'{\i}sica Moderna,
  Universidad de Granada, E-18071 Granada, Spain.}
\author{J.M. Verde--Velasco} \affiliation{Grupo de F\'\i sica Nuclear, 
  Facultad de Ciencias, E-37008 Salamanca, Spain.}


\begin{abstract} 
 The semileptonic $B\to \pi l^+ \nu_l$ decay is studied starting from
 a simple quark model which includes the influence of the $B^*$ pole.
 To extend the predictions of a nonrelativistic constituent quark
 model from its region of applicability near $q^2_{\rm
   max}=(m_B-m_\pi)^2$ to all $q^2$ values accessible in the physical
 decay, we use a novel multiply-subtracted Omn\`es dispersion
 relation, which considerably diminishes the form factor dependence on
 the elastic $\pi B \to \pi B$ scattering amplitudes at high energies.
 By comparison to the experimental branching fraction we extract
 $|V_{ub}| = 0.0034 \pm 0.0003\, ({\rm exp}) \pm 0.0007\, ({\rm
   theory})$. To further test our framework, we also study $D\to \pi$
 and $D\to K$ decays and find excellent results
 $\frac{f^+_\pi(0)}{f^+_K(0)} = 0.80 \pm 0.03, \quad \frac{{\cal
     B}(D^0\to \pi^- e^+ \nu_e)}{{\cal B}(D^0\to K^- e^+ \nu_e)}=
 0.079 \pm 0.008$. In particular for the $D\to \pi$ case, we
 reproduce, with high accuracy, the three-flavor lattice QCD results
 recently obtained by the Fermilab-MILC-HPQCD Collaboration. While for
 the $D\to K$ case, we successfully describe the data for
 $f^+(q^2)/f^+(0)$ recently measured by the FOCUS Collaboration.
\end{abstract}
\maketitle

\section{Introduction}

Exclusive semileptonic decays of $B$-mesons are of great interest,
since they can be used to determine the Cabibbo-Kobayashi-Maskawa
(CKM) matrix elements $|V_{ub}|$ and $|V_{cb}|$. In the latter case,
heavy quark symmetry greatly simplifies the theoretical understanding
of the hadronic transition matrix elements and thus the overall
theoretical uncertainty on the decay process is under
control~\cite{pdg}. The measurement of the exclusive semileptonic
decay $B\to \pi l^+ \nu_l$ by the CLEO
Collaboration~\cite{Exp_96,Exp_03} can be used to determine the CKM
parameter $|V_{ub}|$. This exclusive method provides an important
alternative to the extraction of $|V_{ub}|$ from inclusive
measurements of $B\to X_u l^+ \nu_l$. For semileptonic decays of
charmed or bottom mesons into light mesons there are no flavor
symmetries to constrain the hadronic matrix elements, and as a result,
the errors on $|V_{ub}|$ are currently dominated\footnote{ The current
  best value for $|V_{cd}|$ comes from neutrino production of charm
  off valence $d$ quarks (with the cross section from perturbative
  QCD), rather than from semileptonic $D$ decays.} by theoretical
uncertainties~\cite{pdg}. An accurate determination of $|V_{ub}|$ with
well--understood uncertainties remains one of the fundamental
priorities for heavy flavor physics.

The transition amplitude for the exclusive semileptonic $b\to u$
decays factorizes into leptonic and hadronic parts. The hadronic
matrix elements contain the non-perturbative, strong--interaction
effects and have been extensively evaluated within different
approaches. Thus, several lattice QCD (first in the quenched
approximation, \cite{Latt_96,Latt_98,Latt_98_nrqcd,Latt_00,Latt_01,
Latt_01_bis}, and more recently using dynamical
configurations~\cite{Latt_04_HPQCD,Latt_04_FNLAB}), light-cone sum
rule (LCSR)~\cite{LCSR_98,LCSR_00,LCSR_Hqet_01,LCSR_02,LCSR_03} and
constituent quark model~\cite{QM_85,QM_89,QM_90_IW,QM_90,QM_95,QM_96,
QM_97,QM_00,QM_02} calculations have been carried out in recent
years. Each of the above methods has only a limited range of
applicability, namely: LCSR are suitable for describing the low
squared momentum transfer ($q^2$) region of the form factors, while
Lattice QCD, because of the limitation on the magnitude of spatial
momentum components, provides results only for the high $q^2$
region. Constituent quark models may give the form factors in the full
$q^2$ range, but they are not closely related to the QCD
Lagrangian\footnote{A rigorous derivation of this approach as an
effective theory of QCD in the non--perturbative regime has not been
obtained.} and therefore have input parameters which are not directly
measurable and might not be of fundamental significance.  Thus, it is
evident that a combination of various methods is required.

Watson's theorem for the $B\to \pi l^+ \nu_l$ process allows one to
write a dispersion relation for each of the form factors entering in
the hadronic matrix element. This procedure leads to the so-called
Omn\`es representation~\cite{Omnes}, which can be used to constrain
the $q^2$ dependence of the form factors from the elastic $\pi B \to
\pi B$ scattering amplitudes~\cite{Omnes_01}. In Ref.~\cite{Omnes_01},
once-subtracted dispersion relations were used, and though promising
results were found, they suffered from sizeable uncertainties because
of imprecise knowledge of the $\pi B \to \pi B$ phase shifts far from
threshold. A recent re-analysis of the Omn\`es representation in this
context~\cite{Omnes_05}, has shown that the use of multiply-subtracted
dispersion relations considerably diminishes the form factor
dependence on the elastic $\pi B \to \pi B$ scattering amplitudes at
high energies, and more importantly points out that the Omn\`es
representation of the form factors can be used to combine predictions
from various methods in different $q^2$ regions.

In this paper we study the semileptonic $B\to \pi l^+ \nu_l$ decay. We
take advantage of the findings of Ref.~\cite{Omnes_05} and use the
predictions of LCSR calculations at $q^2=0$ to extend the predictions
of a simple nonrelativistic constituent quark model (NRCQM) from its
region of applicability (near $q^2_{\rm max}=(m_B-m_\pi)^2$) to all
$q^2$ values accessible in the physical decay. We also use the
available lattice QCD data to test our approach. We use a Monte Carlo
procedure to find theoretical error bands for the form factors and the
decay width. From our estimate of the decay width and the branching
ratio measurement of Ref.~\cite{Exp_03} we obtain
\begin{equation}
|V_{ub}|_{\rm this ~ work} = 0.0034 \pm 0.0003\, ({\rm exp}) \pm
 0.0007\, ({\rm theory}) 
\end{equation}

To further test this simple framework, we also study the $D\to \pi$
and $D\to K$ decays, for which there exist precise experimental data
and for which the relevant CKM matrix elements ($|V_{cd}|$ and
$|V_{cs}|$) are also well known. We find
\begin{equation}
f^+_{\pi}(0)  = 0.63 \pm 0.02,\quad f^+_{ K}(0)  = 0.79 \pm 0.01, \quad
\frac{f^+_{\pi}(0)}{f^+_{K}(0)} = 0.80 \pm 0.03, \quad 
\frac{{\cal B}(D^0\to \pi^- e^+ \nu_e)}{{\cal B}(D^0\to
  K^- e^+ \nu_e)}\Big|_{\rm this~work}= 0.079 \pm 0.008
\end{equation}

The plan of this paper is as follows. After this introduction, we
study the semileptonic $B\to \pi$ decay in Sect.~\ref{sec:b2pi}. First
we set up the form factor decomposition (Subsect.~\ref{sec:b2pia}),
discuss the valence quark approximation (Subsect.~\ref{sec:val}) and
the role played by the $B^*$ resonance (Subsect.~\ref{sec:pole}). The
Omn\`es dispersion relation and its application to this decay is
addressed in Subsect.~\ref{sec:omnes} and in the Appendix. Finally, in
Subsect.~\ref{sec:err} we use our framework to determine $|V_{ub}|$,
paying special attention to estimating the uncertainties of the
determination. In Sect.~\ref{sec:dpik} we study the $D\to \pi$ and
$D\to K$ semileptonic decays and finally in Sect.~\ref{sec:concl} we
present our conclusions.

\section{Semileptonic $B\to \pi$ Decays}

\label{sec:b2pi}

\subsection{Differential Decay Width and Form Factor Decomposition}
\label{sec:b2pia}

Using Lorentz, parity, and time-reversal invariance, the matrix
element for the semileptonic $B^0\to \pi^- l^+ \nu_l$ decay can be
parametrized in terms of two invariant and dimensionless form factors
as\footnote{Note that the axial current does not contribute to
transitions between pseudoscalar mesons.}
\begin{equation}
\langle \pi (p_\pi) | V^\mu | B (p_B) \rangle =
\left(p_B+p_\pi-q\frac{m_B^2-m_\pi^2}{q^2} \right)^\mu f^+(q^2) +
q^\mu \frac{m_B^2-m_\pi^2}{q^2} f^0(q^2) 
\end{equation}
where $q^\mu = (p_B-p_\pi)^\mu$ is the four momentum transfer and
$m_B=5279.4$ MeV and $m_\pi=139.57$ MeV are the $B^0$ and $\pi^-$
masses, respectively. The physical meaning of the form factors is
clear in the helicity basis, in which $f^+$ ($f^0$) corresponds to a
transition amplitude with $1^-$ ($0^+$) spin--parity quantum numbers
in the center of mass of the lepton pair. For massless leptons ($l=e$
or $\mu$), the total decay rate is given by
\begin{equation}
\Gamma\left(B^0\to \pi^- l^+ \nu_l \right) =
\frac{G_F^2|V_{ub}|^2}{192\pi^3m^3_B} \int_0^{q^2_{\rm max}}
dq^2\left[\lambda (q^2)\right]^\frac32 |f^+(q^2)|^2 \label{eq:gamma}
\end{equation}
with $q^2_{\rm max}=(m_B-m_\pi)^2$, $G_F= 1.16637\times 10^{-5}\gev^{-2}$ and
$\lambda(q^2)=(m^2_B+m^2_\pi-q^2)^2-4m^2_Bm^2_\pi=4m^2_B|\vec{p}_\pi|^2$,
with $\vec{p}_\pi$ the pion three-momentum in the $B$ rest frame.

Measurements of the $B^0$ lifetime, $\tau_{B^0}= (1.536\pm 0.014)
\times 10^{-12}\,\mathrm{s}$ and of the $B^0\to \pi^- l^+ \nu_l$
branching fraction, ${\cal B}_{\rm exp}(B^0\to \pi^- l^+ \nu_l)=(1.33
\pm 0.22)\times 10^{-4}$~\cite{pdg} lead to
\begin{equation}
\Gamma_{\rm exp}\left(B^0\to \pi^- l^+ \nu_l \right) = (8.7\pm 1.5) \times
10^7~ {\rm s}^{-1} = (5.7 \pm 1.0) \times 10^{-14} ~{\rm MeV}, \quad
l=e~ {\rm or}~ \mu \label{eq:exp}
\end{equation}

\subsection{ Nonrelativistic Constituent Quark Model: Valence Quark
  Contribution}
\label{sec:val}

Within the spectator approximation, considering only the valence quark
contribution and assuming that the $B$ and $\pi$ mesons are $S$-wave
quark-antiquark bound states, a NRCQM (with constituent quark masses
$m_b$ and $m_l=m_u=m_d$) predicts~\cite{NRCQM_def}:
\begin{eqnarray}
\frac{\left\langle \pi (E_\pi,-\vec{q}\ ) \Big| V^\mu \Big| B (m_B,\vec{0}\ )
  \right\rangle^{\rm val}}{\sqrt{4m_BE_\pi}} &=& \int \frac{d^3l}{4\pi}
  \sqrt{\frac{E_b(\vec{l}\, )+m_b}{2E_b(\vec{l}\,)}}
  \sqrt{\frac{E_u(\vec{l}+\vec{q}\, )+m_u}{2E_u(\vec{l}+\vec{q}\, )}}
  \phi^B_{\rm rel}(|\vec{l}\,|)\phi^\pi_{\rm rel}(|\vec{l}+
\frac{m_{sp}}{m_u+m_{sp}}\vec{q}\,|) {\cal
  V}^\mu(\vec{l},\vec{q}\,)\nonumber\\
&&\nonumber\\
{\cal
  V}^\mu(\vec{l},\vec{q}\,)&=&\left(\begin{array}{c}1+\frac{\vec{l}^{\,\,2}+
  \vec{l}\cdot\vec{q}}{(E_b(\vec{l}\,)+m_b)(E_u(\vec{l}+\vec{q}\,)+m_u)}\\
  \\-\frac{\vec{l}}{E_b(\vec{l}\,)+m_b}-\frac{\vec{l}+\vec{q}}{E_u(\vec{l}+
  \vec{q}\,)+m_u)
}\end{array} \right) \label{eq:vmu}
\end{eqnarray}
with $E_\pi=\sqrt{m_\pi^2+\vec{q}^{\,\,2}}$, $E_{b,u}(\vec{k})=
\sqrt{m_{b,u}^2+\vec{k}^{\,2}}$, $m_{sp}$ the spectator quark mass
($m_d$ in this case) and $\phi^{B,\pi}_{\rm rel}(k)$ the Fourier
transforms of the radial coordinate space $B,\pi$ meson wave
functions, which describe the relative dynamics of the
quark-antiquark pair\footnote{They are normalized to
  $\int_0^{+\infty}dk k^2|\phi^{B,\pi}_{\rm rel}(k)|^2=1$}.

To evaluate the coordinate space wave function
we have used several nonrelativistic  quark-antiquark
interactions. Their general structure is as follows~\cite{QM_pot_1,QM_pot_2}
\begin{equation}
V_{ij}^{q\bar q}(r) = -\frac{\kappa\left ( 1 - e^{-r/r_c}\right )}{r}
+\lambda r^p - \Lambda + \left\{a_0\frac{\kappa}{m_im_j}
\frac{e^{-r/r_0}}{rr_0^2} + \frac{2\pi}{3m_im_j}\kappa^\prime \left (
1 - e^{-r/r_c}\right ) \frac{e^{-r^2/x_0^2}}{\pi^\frac32 x_0^3} \right
\}\vec{\sigma}_i\vec{\sigma}_j  \label{eq:phe}
\end{equation}
with $\vec{\sigma}$ the spin Pauli matrices, $m_i$ the constituent
quark masses  and
\begin{equation}
x_0(m_i,m_j) = A \left ( \frac{2m_im_j}{m_i+m_j} \right )^{-B} 
\label{eq:phebis}
\end{equation}
The potentials considered differ in the form factors used for the
hyperfine terms, the power of the confining term ($p=1$, as suggested
by lattice QCD calculations~\cite{GM84}, or $p=2/3$ which gives the
correct asymptotic Regge trajectories for mesons~\cite{Fabre88}), or
the use of a form factor in the one gluon exchange Coulomb potential.
All interactions have been adjusted to reproduce the light ($\pi$,
$\rho$, $K$, $K^*$, etc.) and heavy-light ($D$, $D^*$, $B$, $B^*$,
etc.) meson spectra and lead to precise predictions for the charmed
and bottom baryon ($\Lambda_{c,b}$, $\Sigma_{c,b}$, $\Sigma^*_{c,b}$,
$\Xi_{c,b}$, $\Xi'_{c,b}$, $\Xi^*_{c,b}$, $\Omega_{c,b}$ and
$\Omega_{c,b}^*$) masses~\cite{QM_pot_2,QM_hqs} and for the
semileptonic $\Lambda_b^0 \to \Lambda_c^+ l^- {\bar \nu}_l$ and
$\Xi_b^0 \to \Xi_c^+ l^- {\bar \nu}_l$~\cite{QM_slp} decays.

\begin{figure}
\begin{center}
\makebox[0pt]{\input{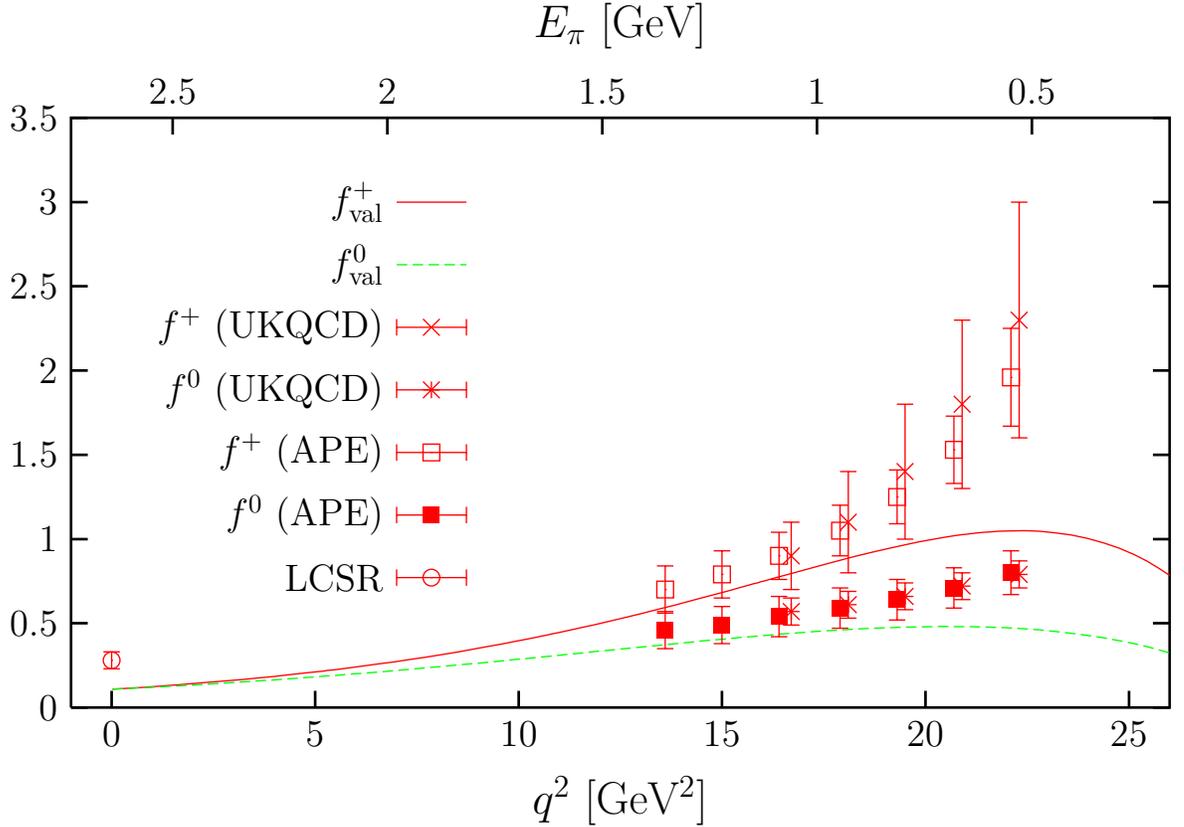}}
\end{center}
\caption{NRCQM valence quark $f^+$ and $f^0$ form factors from the AL1
inter-quark interaction~\protect\cite{QM_pot_2}. Lattice data points are
taken from Refs.~\protect\cite{Latt_00} (UKQCD)
and~\protect\cite{Latt_01} (APE) and the LCSR estimate at $q^2=0$ is
given in Ref.~\protect\cite{LCSR_00}.}\label{fig:valence}
\end{figure}

Typical NRCQM valence quark predictions for the $f^+$ and $f^0$ form
factors are depicted in Fig.~\ref{fig:valence}. The AL1 potential from
Ref.~\cite{QM_pot_2} has been used\footnote{The sensitivity of the
  results to the quark-antiquark nonrelativistic interaction will be
  discussed in detail later.} and for comparison quenched lattice
results are also plotted. Preliminary unquenched lattice calculations
have been presented recently~\cite{Latt_04_HPQCD,Latt_04_FNLAB}, but
no significant difference between quenched and unquenched calculations
is observed~\cite{Has04}, within relatively large statistical errors.
In addition, LCSR provide accurate and theoretically well founded
results in the $q^2=0$ region. Thus, we have a LCSR
value~\cite{LCSR_00}
\begin{equation}
{\rm LCSR:}~ f^+(0) = 0.28 \pm 0.05
\label{eq:lcsr}
\end{equation}
which is also plotted in Fig.~\ref{fig:valence}.  

Fig.~\ref{fig:valence} clearly shows the deficiencies of the NRCQM
valence quark description of the $B\to \pi l^+ \nu_l$ semileptonic
decay. It fails over the full range of $q^2$ values. Close to
$q^2_{\rm max}$, where the nonrelativistic Schr\"odinger equation
should work best, the influence of the $B^*$ resonance is clearly
visible~\cite{QM_90_IW}. At the opposite end, close to $q^2=0$, where
$|\vec{q}\,|\approx 2.5\gev$, predictions from a nonrelativistic
scheme are clearly not trustworthy. As a result, a value for the width
$\Gamma_{\rm NRCQM}^{\rm val}\left(B^0\to \pi^- l^+ \nu_l \right) =
2.4 \left(\frac{|V_{ub}|}{0.0032}\right)^2 \times 10^{-14}\mev$ is
obtained, which is around a factor of two smaller than the CLEO
measurement quoted in Eq.~(\ref{eq:exp}).

\subsection{ Nonrelativistic Constituent Quark Model: $B^*$ Resonance
  Contribution}

\label{sec:pole}

\begin{figure}[tbh]
\centerline{\includegraphics[height=3cm]{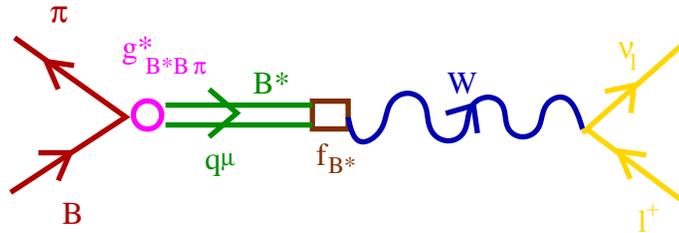}}
\caption{$B^*$ resonance contribution to the $f^+$ form factor for the
  semileptonic $B \to \pi$ decay.}
\label{fig:bstar}
\end{figure}

A NRCQM description of the decay process should be feasible in the
neighborhood of $q^2_{\rm max}$. Indeed, this is the case for the
semileptonic $B\to D l {\bar \nu}_l$ and $B\to D^* l {\bar \nu}_l$
decays, recently studied in Ref.~\cite{NRCQM_def} with the same NRCQM
as here. The difference here is that, as first pointed out in
Ref.~\cite{QM_90_IW}, in the chiral limit and as $m_b \to \infty$, the
decay $B^0\to \pi^- l^+ \nu_l$ should be dominated near zero pion
recoil by the effects of the $B^*$ resonance, which is quite close to
$q^2_{\rm max}$. In the picture of Ref.~\cite{QM_90_IW}, the one we
will adopt here, the $B^*$ contribution plays a role only near
$q^2_{\rm max}$, since it is strongly suppressed by a soft hadronic
vertex. This is in sharp contrast to phenomenological
parameterizations of $f^+$ which assume it dominates over the full
range accessible in the physical decay~\cite{Latt_96}. The $B^*$
effects of the type considered here are not dual to the valence quark
model form factors and must be added as a distinct coherent
contribution to heavy quark decay near $q^2_{\rm
  max}$~\cite{QM_90_IW}. We will focus on the $f^+$ form factor, which
determines the decay width for massless leptons, and we evaluate the
contribution to it from the diagram depicted in Fig.~\ref{fig:bstar}.
It leads to a hadronic amplitude (normalizations as in
Ref.~\cite{NRCQM_def})
\begin{equation}
-i T^\mu = - i \widehat{g}_{B^*B\pi} (q^2) p_\pi^\nu  \left ( i
 \frac{-g_{\nu}^\mu+q^\mu q_\nu/m_{B^*}^2}{q^2-m_{B^*}^2}\right ) i
 \sqrt{q^2} \widehat{f}_{B^*}(q^2) \label{eq:tbstar}
\end{equation}
with $m_{B^*}=5325$ MeV, and $\widehat{f}_{B^*}$ and
$\widehat{g}_{B^*B\pi}$ the $B^*$ decay constant and the strong
$B^*B\pi$ dimensionless coupling constant for a virtual $B^*$ meson,
respectively. On the $B^*$ mass shell, the hadron matrix elements
$\widehat{f}_{B^*}(q^2=m_{B^*}^2)\equiv f_{B^*}$ and
$\widehat{g}_{B^*B\pi}(q^2=m_{B^*}^2)\equiv g_{B^*B\pi} $ reduce to
the ordinary $B^*$ decay constant and coupling of a pion to $B$ and
$B^*$ mesons. The latter is related, in the heavy quark limit, to
$\widehat{g}$, the coupling of the vector and pseudoscalar heavy-light
mesons to the pion~\cite{defg,hmchpt}\footnote{We use the
  normalization $f_\pi \approx 131\mev$.}
\begin{equation}
g_{B^*B\pi}=\left(\frac{2\widehat{g}\sqrt{m_Bm_{B^*}}}{f_\pi}\right)(1+
{\cal O}(1/m_b))\label{eq:ghat}
\end{equation}

From  Eq.~(\ref{eq:tbstar}) we get
\begin{equation}
f^+_{\rm pole}(q^2) = \frac12 \widehat{g}_{B^*B\pi} (q^2) \frac{\sqrt{q^2}
  \widehat{f}_{B^*}(q^2)}{m_{B^*}^2-q^2} \label{eq:polefg}
\end{equation}

There is no direct experimental determination of $g_{B^*B\pi}$,
because there is no phase space for the decay $B^* \to B \pi$. The
available experimental results for $D^* \to D \pi$~\cite{pdg} can be
related to $g_{B^*B\pi}$, through heavy quark symmetry. There is no
direct measurements of $f_{B^*}$ either.  In Ref.~\cite{NRCQM_def} we
computed, within the same NRCQM approach as the one outlined here, both
$g_{B^*B\pi}$ and $f_{B^*}$, and we found a value of $9.1 \pm 0.9\gev$ for the
product of both quantities, which appears in $f^+_{\rm pole}$ at
$q^2=m_{B^*}^2$. Lattice QCD simulations  have measured
$f_{B^*}$~\cite{Latt_fbstar} and $g_{B^*B\pi}$~\cite{g_Latt_03} to be
\begin{eqnarray}
f_{B^*} = 190 \pm 30\mev \quad 
g_{B^*B\pi} = 47 \pm 5 \pm 8 \Rightarrow
\left[g_{B^*B\pi}f_{B^*}\right]_{\rm Latt-QCD} = 8.9 \pm 2.2\gev
\label{eq:ukqcd_err}
\end{eqnarray}
where we have added errors in quadrature. Thus the lattice prediction
for the product $g_{B^*B\pi}f_{B^*}$ is in remarkable agreement,
within $3\%$, with our NRCQM estimate in~\cite{NRCQM_def}. In what
follows we will use the value and error for the product estimated from
the lattice data and use the NRCQM of Ref.~\cite{NRCQM_def} to
determine the $q^2$ dependence of $\widehat{g}_{B^*B\pi}(q^2)$ and
$\widehat{f}_{B^*}(q^2)$, as we will discuss below. There are other
recent estimates for $g_{B^*B\pi}$ (\cite{g_LCSR_95,g_QM_99}) and
$f_{B^*}$ (\cite{Be02}), but given the existing uncertainties, all of
them are compatible with the lattice values quoted in
Eq.~(\ref{eq:ukqcd_err}).

\begin{figure}
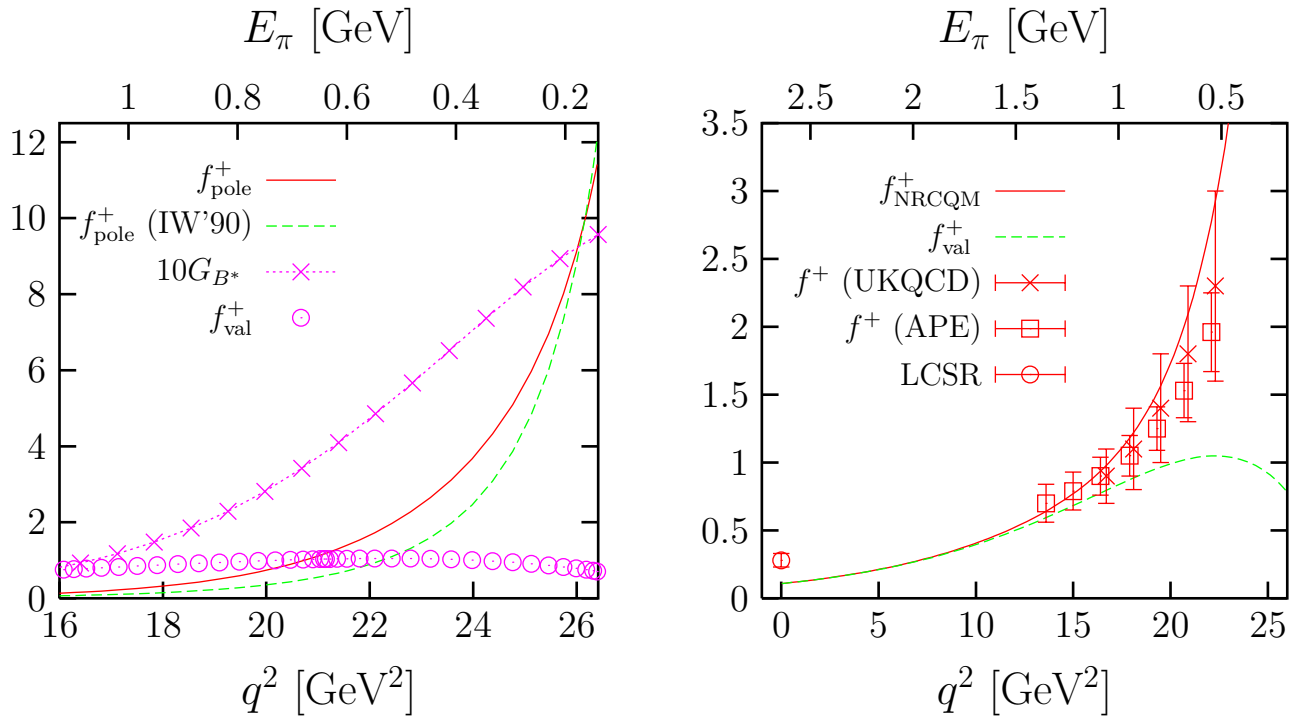

\begin{center}
\makebox[0pt]{\input{pole.tex}\hspace{-10mm}\input{pole-val.tex}}
\end{center}
\caption{ {\bf Left:} The solid line denotes the AL1 NRCQM $B^*$
  pole contribution to $f^+$ (Eq.~(\protect\ref{eq:pol_fi})), while
  the dashed line stands for the $B^*$ contribution to $f^+$ obtained
  within the gaussian constituent quark model of
  Refs.~\protect\cite{QM_89,QM_90_IW}. We also plot the $G_{B^*}$
  hadron factor introduced in Eq.~(\protect\ref{eq:pol_fi}) and the
  valence quark contribution to $f^+$ depicted in
  Fig.~\protect\ref{fig:valence}. {\bf Right:} Valence quark and
  valence quark plus $B^*-$pole  (denoted as NRCQM)  contributions to
  $f^+$. We also plot lattice QCD and LCSR $f^+$ data from the same
  references as in Fig.~\protect\ref{fig:valence}.  }
\label{fig:pol-val}
\end{figure}

As mentioned above, we use the NRCQM framework to estimate the $q^2$
dependence of the product of $\widehat{f}_{B^*}(q^2)
\widehat{g}_{B^*B\pi}(q^2)$. Since the NRCQM always uses on-shell meson wave
functions, all $q^2$ dependence will arise from the kinematical
factors relating the quark model matrix elements and the hadron form
factors. For instance, from Eq.~(13) of Ref.~\cite{NRCQM_def}
we find a rather mild $q^2$ dependence 
\begin{equation}
\widehat{f}_{B^*}(q^2)\sqrt[4]{q^2} = f_{B^*} \sqrt{m_{B^*}}
\end{equation}
In the same manner, we use Eqs.~(50) and (51) of Ref.~\cite{NRCQM_def}
to determine the $q^2$ dependence of $\widehat{g}_{B^*B\pi}$, setting
the $B$-meson four momentum $P'^\mu=(m_B,\vec{P}' = 0)$ and the
$B^*$-meson four momentum $P^\mu=q^\mu=(m_B-E_\pi, \vec{P} =
-\vec{q})$, and off--shell mass given by $\sqrt{q^2}$, 
$|\vec{q}\,|= \sqrt{E_\pi^2-m_\pi^2}$, and $E_\pi$ determined
from $q^2$ as usual ($E_\pi = (m_B^2+m_\pi^2-q^2)/2m_B$). Thus,
finally we evaluate
\begin{equation}
f^+_{\rm pole}(q^2) = \frac12
  G_{B^*}(q^2)\frac{\sqrt[4]{q^2}}{\sqrt{m_{B^*}}}
\frac{m_{B^*}\left[g_{B^*B\pi}f_{B^*}\right]_{\rm
  Latt-QCD}}{m_{B^*}^2-q^2} \label{eq:pol_fi}
\end{equation}
where $G_{B^*}(q^2)= \widehat{g}_{B^*B\pi}(q^2)/g_{B^*B\pi}$ is a
dimensionless hadronic factor normalized to one at $q^2= m_{B^*}^2$,
which accounts for the $q^2$ dependence of $B\to B^* \pi$
amplitude. In Fig.~\protect\ref{fig:pol-val}, we show the influence of
the $B^*$ resonance within our NRCQM and compare our results to those
obtained by Isgur and Wise from the gaussian constituent quark model
of Refs.~\protect\cite{QM_89,QM_90_IW}. Our model for the $B^*$
contribution compares well to that of Ref.~\cite{QM_90_IW}, though the
latter decreases faster owing to the use of a harmonic oscillator
basis. The inclusion of $f^+_{\rm pole}$ clearly improves the simple
valence quark contribution and leads to a reasonable description of
the lattice data from $q^2_{\rm max}$ down to $q^2$ values around
$15\gev^2$. The low $q^2$ region is still poorly described within the
current model since relativistic corrections there should be large.

The hadronic amplitude of Eq.~(\ref{eq:tbstar}) also leads to a small
contribution to the $f^0$ form factor. Though it also improves the
description for the highest $q^2$ values, it is not large enough and
it is necessary to consider the influence of the lightest $0^+$
$B$-resonances~\cite{Omnes_01} (for instance a resonance around
$5660\mev$~\cite{Ci99}).

\subsection{Omn\`es Representation}

\label{sec:omnes}

Here we use the Omn\`es representation of the $f^+$ form factor to
combine the NRCQM predictions at high $q^2$ values, say above
$18\gev^2$, with the LCSR result at $q^2=0$. In this way we obtain the
full $q^2$ dependence of the form factor and thus can determine the
$|V_{ub}|$ CKM matrix element from the integrated semileptonic width.
As shown in the Appendix, the $(n+1)$-subtracted Omn\`es
representation for $f^+$ reads:
\begin{eqnarray}
  f^+(q^2) &=& \bigg( \prod_{j=0}^n
  \left[f^+(q^2_j)\right]^{\alpha_j(q^2)} \bigg)
  \exp\bigg\{{\cal I_\delta}(q^2;\,q^2_0,q^2_1,\cdots,q^2_n)
  \prod_{k=0}^n(q^2-q^2_k)\bigg\}, \nonumber \\ 
{\cal I_\delta}(q^2;\, q^2_0,\cdots,q^2_n) &=&
  \frac{1}{\pi}\int_{\sth}^{+\infty}
  \frac{ds}{(s-q^2_0)\cdots(s-q^2_n)}\frac{\delta(s)}{s-q^2},\nonumber \\
  \alpha_j(q^2) &=& \prod_{j\ne k=0}^{n} \frac{q^2-q^2_k}{q^2_j-q^2_k},
  \label{eq:omnes}
\end{eqnarray}
with $q^2 < \sth=(m_B+m_\pi)^2$ and $q^2_0,\cdots,q^2_n \in
\left]-\infty,\sth\right[$. This representation requires as an input
    the elastic $\pi B \to \pi B$ phase shift $\delta(s)$ in the
    $J^P=1^-$ and isospin $I= 1/2 $ channel plus the form-factor at
    $(n+1)$ $q^2$-values ($q^2_0,q^2_1,\cdots,q^2_n$) below the $\pi
    B$ threshold.

We would like to stress that from a theoretical point of view the
Omn\`es representation is derived from first principles: the
well-established Mandelstam hypothesis~\cite{Ma58} of maximum
analyticity and Watson's theorem~\cite{Wa55}
\begin{eqnarray}
{ f^+(s+ \mi  \epsilon ) \over f^+(s- \mi  \epsilon ) } =
{ T (s+ \mi  \epsilon ) \over T (s- \mi  \epsilon )} =
e^{ 2 \mi  \delta (s) },\,\,\, s > \sth, \qquad 
T(s) = \frac{8\pi \mi  s}{\lambda^\frac12(s)}\left (
e^{2i\delta(s)}-1\right )
\label{eq:wat}
\end{eqnarray}
Omn\`es ideas have been used successfully to account for final state
interactions in kaon decays~\cite{Omnes_01_kaon}\footnote{There,
  however, multiple derivatives evaluated at a single point are used
  as input instead of subtractions for different $q^2$ values.} and in
Ref.~\cite{Omnes_01}, a once-subtracted Omn\`es representation
(subtraction point $q^2_0=0$) was applied to the study of
semi-leptonic $B\to \pi$ decays. In the latter work phase shifts were
evaluated by solving the Bethe-Salpeter equation in the so-called
\emph{on-shell scheme}~\cite{EJ99}, with a kernel determined by the
direct tree level amplitude from the lowest order heavy meson chiral
perturbation theory lagrangian~\cite{hmchpt}, together with the tree
diagrams for $B^*$ exchange which involve the leading interaction with
coupling $\widehat{g}$. Such a model accommodates the $B^*$ as a $\pi
B$ bound state and should acceptably describe phase shifts close to
threshold. It led to promising results for $f^+$~\cite{Omnes_01}, but
theoretical uncertainties on the form factor were not negligible,
since to compute the Omn\`es factor ${\cal I}_\delta$
(Eq.~(\ref{eq:omnes})) requires elastic phase-shifts far from
threshold\footnote{Higher resonance effects on phase shifts cannot be
  neglected far from threshold. In particular the LCSR result at
  $q^2=0$ hints that at least an extra $J^P=1^-$ resonance, located
  around $6\gev$, has to be included in the once-subtracted Omn\`es
  relation scheme~\protect\cite{Omnes_01}.}.  To include the effects of
higher resonances on $\delta(s)$ requires input of the masses
and couplings of such resonances. We therefore make many subtractions
in the Omn\`es dispersion relation to suppress the impact of
$\delta(s)$ at large $s$. This will leave a systematic effect in our
results, but this should be less than that coming from the NRCQM plus
$B^*$ pole used as our main input.

\begin{figure}
\begin{center}
\makebox[0pt]{\input{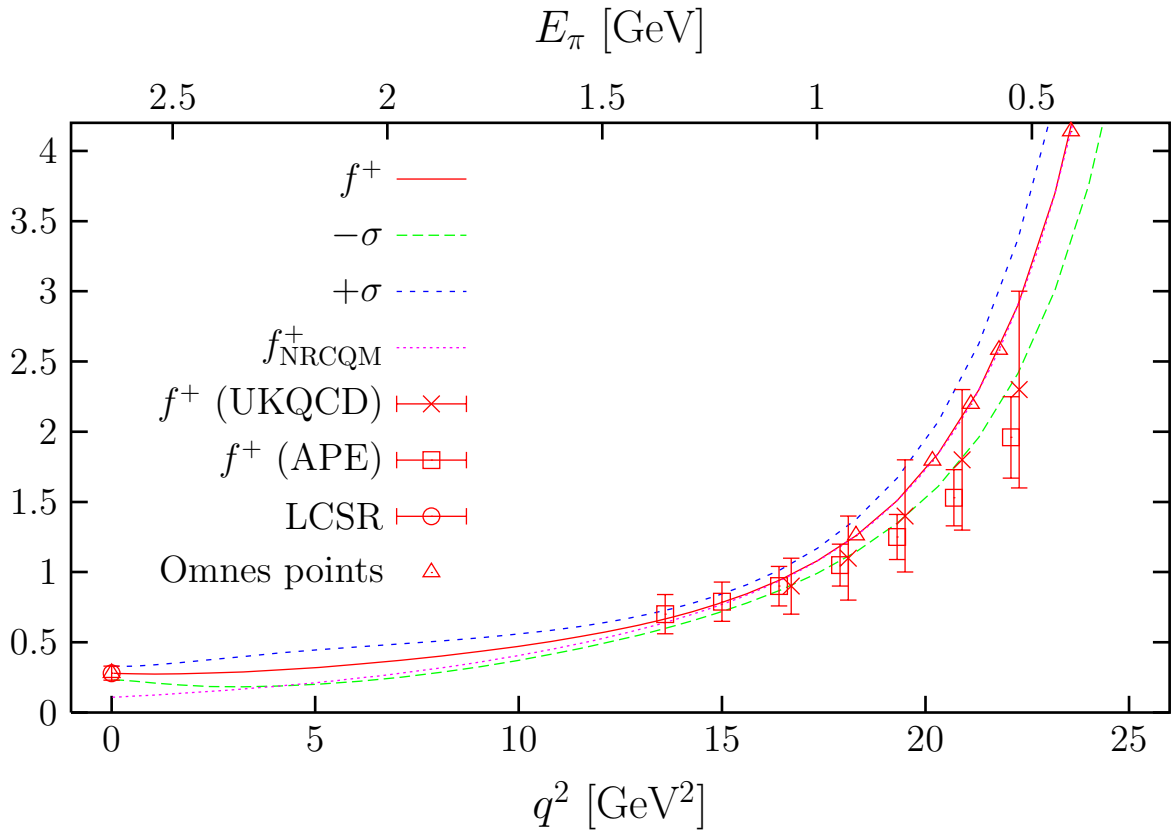}}
\end{center}
\caption{Solid line: $f^+$ form factor from a $6$-subtracted Omn\`es
  representation, Eq.~(\ref{eq:omn_th}). Triangles denote the input
  subtraction points of the Omn\`es dispersion relation (five points
  obtained from our AL1 NRCQM plus the Ref.~\cite{LCSR_00} LCSR result
  at $q^2=0$). Lattice data are from Refs.~\cite{Latt_00}(UKQCD)
  and~\cite{Latt_01}(APE), and the AL1 valence quark plus $B^*$-pole
  contribution to $f^+$ is also shown (denoted $f^+_\mathrm{NRCQM}$).
  Finally, $\pm \sigma$ lines show the theoretical uncertainty bands
  on the Omn\`es form factor inherited from the errors in
  Eq.~(\ref{eq:ukqcd_err}) and from the quark-antiquark interaction
  model dependence (see Subsect.\ref{sec:err} for details). }
\label{fig:omnes}
\end{figure}

As the number of subtractions increases the integration region
relevant in Eq.~(\ref{eq:omnes}) gets reduced and, if this number is
large enough, only the phase shifts at or near threshold will be
needed. Note that close to threshold the $p$-wave
phase shift behaves as
\begin{equation}
\delta(s) = n_b \pi - p^3 a + \cdots \label{eq:levin}
\end{equation}
where $n_b$ is the number of bound states in the channel (Levinson's
theorem~\cite{MS70}), $p$ is the $\pi B$ center of mass momentum and
$a$ the corresponding scattering volume. In our case $n_b=1$ if we
consider the $B^*$ as a $\pi B$ bound state. Here, we will perform a
large number of subtractions so that approximating $\delta(s)\approx
\pi$ in Eq.~(\ref{eq:omnes}) will be justified. The Omn\`es factor
${\cal I}_\delta$ can then be evaluated analytically and we find for
$q^2 < \sth$
\begin{eqnarray}
  f^+(q^2) &\approx& \frac{1}{\sth -q^2} \prod_{j=0}^n
  \left[f^+(q^2_j)(\sth-q^2_j)\right]^{\alpha_j(q^2)}, \qquad n \gg 1
  \label{eq:omn_th}
\end{eqnarray}
Next we use the above formula to combine the LCSR result at $q^2=0$
and those obtained from our NRCQM in the high $q^2$ region and
presented in the previous section. Thus we have used the $f^+$ NRCQM
(valence $+$ pole) predictions for five $q^2$ values ranging from
$q^2_{\rm max}$ down\footnote{From Eq.~(\protect\ref{eq:vmu}), we see
  that the arguments of the meson wave function are $|\vec{l}|$ and
  $|\vec{l}+\vec{q}/2|$. For $q^2=18\gev^2$, half of the transferred
  momentum, $|\vec{q}\,|/2$, is about $0.4$--$0.5\gev$, which is of
  the same order as $\langle \vec{l}^{\,2}
  \rangle^\frac12_{\phi^{B,\pi}}$. Since the non-relativistic
  quark-antiquark interactions, $V(r)$, have been adjusted to
  reproduce the meson binding energies, they effectively incorporate
  some relativistic corrections and hence one might expect this
  effective nonrelativistic framework to provide reasonable meson wave
  functions for momenta of order $\langle \vec{l}^{\,2}
  \rangle^\frac12_{\phi^{B,\pi}}$. This could explain why the NRCQM
  describes the lattice data (right panel of
  Fig.~\protect\ref{fig:pol-val}) from high values of $q^2$ down to
  values of $q^2$ even smaller than $18\gev^2$. Nevertheless, we find
  it surprising that the nonrelativistic constituent quark model works
  as well as it does~\protect\cite{QM_95}.} to about $18\gev^2$:
\begin{equation}
(q^2\!/\!\gev^2,f^+(q^2)) = 
\left\{ \begin{array}{l}
(23.574,4.1373),\\
(21.804,2.5821),\\
(21.116,2.1969),\\
(20.173,1.7916),\\
(18.290,1.2591)
\end{array}\right.
\end{equation}
together with the LCSR result of
Eq.~(\ref{eq:lcsr}) at $q^2=0$. When one uses a large number of
subtractions, as is the case here, the $\alpha_j$ exponents become
large and there are huge cancellations (note the normalization
condition given in Eq.~(\ref{eq:alphas})). This is the reason why
above, and to ensure numerical stability, we have quoted five
significant digits for the NRCQM input. We are aware that
uncertainties are larger than a precision of five digits, and we will
carefully take this fact into account below. Results are shown in
Fig.~\ref{fig:omnes}. As can be seen there, we obtain a simultaneous
description of both lattice data in the high $q^2$ region and the LCSR
prediction at $q^2=0$. In this way, starting from a nonrelativistic
valence quark picture of the semileptonic process
(Subsect.~\ref{sec:val}) with all its obvious limitations, we have
ended up with a realistic description of the relevant form factor for
all $q^2$ values accessible in the physical decay.

A final remark concerns the use of the simplified Omn\`es
representation of Eq.~(\ref{eq:omn_th}) instead of the exact one of
Eq.~(\ref{eq:omnes}). For instance, if we use five subtractions (we
drop the NRCQM point at $q^2=21.1\gev^2$) and the full Omn\`es
representation\footnote{We use the model of
  Ref.~\protect\cite{Omnes_01} to obtain the phase-shifts.} of
Eq.~(\ref{eq:omnes}), we find tiny differences from the results shown
in Fig.~\ref{fig:omnes}. These differences are negligible (below
$1\%$) above $10\gev^2$, and though larger, still quite small (around
$5$-$7\%$ at most in the $5\gev^2$ region) below $10\gev^2$.

\subsection{Determination of $|V_{ub}|$: Error Analysis}
\label{sec:err}

The CKM element $|V_{ub}|$ can be determined by comparing the
experimental decay width (Eq.~(\ref{eq:exp})) with the result of
performing the phase space integration of Eq.~(\ref{eq:gamma}) using
the form factor $f^+$ determined in the previous subsection. Here, we
will pay special attention to estimating the theoretical
uncertainties. We have two main sources of theoretical errors:
\begin{enumerate}
\item \emph{Uncertainties in the constituent quark-antiquark
  nonrelativistic interaction:} To estimate those, we will evaluate
  the spread of integrated widths obtained when 
  five different potentials (AL1, AL2, AP1, AP2 and BD, in the
  notation of Ref.~\cite{QM_pot_2}) are considered. The forms and main
  characteristics of those potentials were discussed in Eqs.~(\ref{eq:phe})
  and~(\ref{eq:phebis}). As  mentioned in
  Subsect.~\ref{sec:val}, all interactions have been adjusted to
  reproduce the light and heavy--light meson spectra and lead to
  precise predictions for the charmed and bottom baryon
  masses~\cite{QM_pot_2} and for the semileptonic $\Lambda_b^0 \to
  \Lambda_c^+ l^- {\bar \nu}_l$ and $\Xi_b^0 \to \Xi_c^+ l^- {\bar
  \nu}_l$~\cite{QM_slp} decays.
\item \emph{Uncertainties on $\left[g_{B^*B\pi}f_{B^*}\right]$ and on
  the input to the multiply-subtracted Omn\`es representation:} Errors
  on $\left[g_{B^*B\pi}f_{B^*}\right]$, quoted in
  Eq.~(\ref{eq:ukqcd_err}), affect the $B^*$ pole contribution to
  $f^+$ (see Eq.~(\ref{eq:pol_fi})) and also induce uncertainties in
  the NRCQM prediction for the five points used as input to the
  Omn\`es representation in Eq.~(\ref{eq:omn_th}). Quark-antiquark
  potential uncertainties, discussed in the previous item, also
  induce uncertainties in the Omn\`es input. The errors on
  the $q^2=0$ data point (LCSR), quoted in Eq.~(\ref{eq:lcsr}), should
  also be taken into account.
\end{enumerate}
To take these uncertainties into account, we proceed in two steps:
\begin{enumerate}
\item We fix the quark-antiquark potential to the AL1 interaction as
  in all previous subsections. By means of a Monte Carlo simulation,
  we generate a total of $1000$ $(\left[g_{B^*B\pi}f_{B^*}\right]_{\rm
    Latt-QCD},f^+(0)_{\rm LCSR})$ pairs\footnote{We have checked that
    the errors quoted in the following are already stable when $500$
    event simulations are performed.} from an uncorrelated two
  dimensional gaussian distribution, with central values and standard
  deviations taken from Eqs.~(\ref{eq:ukqcd_err}) and~(\ref{eq:lcsr}),
  respectively. For each of the $1000$ pairs we build up the six
  points that we use in our Omn\`es scheme (Eq.~(\ref{eq:omn_th})) and
  thus find $1000$ different determinations of $f^+$ over the whole
  $q^2$ range accessible in the $B\to \pi$ decay. For each value of
  $q^2$, we discard the highest and lowest $16\%$ of the values
  obtained for the form factor, to leave a $68\%$ confidence level
  band which forms part of the theoretical uncertainty shown in
  Fig.~\ref{fig:omnes}. Since the output distributions are not
  gaussian in general, this accounts for possible skewness.

  Performing the phase space integration for each of the $1000$ form
  factor samples and again discarding the highest and lowest $16\%$ of
  the values, we find
\begin{equation}
\frac{\Gamma\left(B^0\to \pi^- l^+ \nu_l \right)}{|V_{ub}|^2} =
\left (0.50^{+0.14}_{-0.10}\right) \times 10^{-8}\mev
\end{equation}
\item We fix the $(\left[g_{B^*B\pi}f_{B^*}\right]_{\rm
  Latt-QCD},f^+(0)_{\rm LCSR})$ pair to their central values and
  compute the decay width with each of the five quark-antiquark
  interactions discussed above. From the spread of output values, we
  find
\begin{equation}
\frac{\Gamma\left(B^0\to \pi^- l^+ \nu_l \right)}{|V_{ub}|^2} =
\left (0.50\pm 0.15\right) \times 10^{-8}\mev
\end{equation}
\end{enumerate}
Adding both sources of error in quadrature, we get
\begin{equation}
\frac{\Gamma\left(B^0\to \pi^- l^+ \nu_l \right)}{|V_{ub}|^2} =
\left (0.50 \pm 0.20 \right) \times 10^{-8}\mev
\end{equation}
and comparing to the measurement of the width in Eq.~(\ref{eq:exp})
we find
\begin{equation}
|V_{ub}|_{\rm this ~ work} =
 0.0034 \pm 0.0003\, ({\rm exp}) \pm 0.0007\, ({\rm theory}) 
\end{equation}
The CLEO Collaboration~\cite{Exp_03} obtains from studies
of the $B^0\to \pi^- l^+ \nu_l$ branching fraction and $q^2$
distributions, using LCSR for $0\le q^2<16\gev^2$ and lattice QCD
for $16\gev^2 \le q^2 < q^2_{\rm max}$,
\begin{equation}
|V_{ub}|_{\rm CLEO} = 0.0032 \pm 0.0003\, ({\rm exp}) ^{+0.0006}_{-0.0004}\,
 ({\rm theory})
\end{equation}
We see that both determinations of $|V_{ub}|$ are in an excellent
agreement and that in both cases the error is dominated by
uncertainties in the theoretical treatment. We have also calculated
partially integrated branching ratios,
\begin{equation}
{\cal B}(q_1^2\le q^2< q_2^2) = \frac{{\cal B}^{\rm total}_{\rm exp} 
(B^0\to \pi^- l^+ \nu_l)}{\Gamma}  \int_{q_1^2}^{q^2_2} dq^2
\frac{{\rm d}\Gamma}{{\rm d}q^2}\label{eq:bra}
\end{equation}
Theoretical uncertainties partially cancel in the ratio
$\int_{q_1^2}^{q^2_2} dq^2 \frac{{\rm d}\Gamma}{{\rm d}q^2} / \Gamma$.
Our results are compiled in the Table \ref{tab:bra}. There it can
be seen that they compare reasonably well with those quoted in
Ref.~\cite{Exp_03}.  
\begin{table}[tbh]
\begin{center}
\begin{tabular}{c|ccc}
\hline
& \multicolumn{3}{c}{$B^0\to \pi^- l^+ \nu_l$}\\
\hline
\jtstrut
& ${\cal B}(0\le q^2<8\gev^2)/10^{-5}$
 & ${\cal B}(8 \le q^2<16\gev^2)/10^{-5}$
 & ${\cal B}(q^2\ge 16\gev^2)/10^{-5}$  \\[0.5ex]
\hline
\jtstrut
CLEO~\protect\cite{Exp_03}
 & $4.3 \pm 1.2$ & $6.5 \pm 1.3$ & $2.5\pm 1.0$\\
\jtstrut
This work
 & $4.3 \pm 0.7\,[{\rm exp}]\pm 1.2\, [{\rm theory}]$
 & $4.1 \pm 0.7\,[{\rm exp}]\pm 0.4\, [{\rm theory}]$
 & $4.9 \pm 0.8\,[{\rm exp}]\pm 1.2\, [{\rm theory}]$ \\[0.5ex]
\hline
\end{tabular}
\end{center}
\caption{Partially integrated branching ratios (see Eq.~(\ref{eq:bra})).}
\label{tab:bra}
\end{table}

Finally, at each value of $q^2$ we also compute the spread of values
obtained for the $f^+$ form factor when the five different
quark-antiquark interactions are used. This procedure gives us a
further theoretical error on $f^+(q^2)$ at fixed $q^2$ and by adding
it in quadrature to that obtained previously from uncertainties on
the $(\left[g_{B^*B\pi}f_{B^*}\right]_{\rm Latt-QCD},f^+(0)_{\rm LCSR})$
pair, we determine the theoretical error bands shown in
Fig.~\ref{fig:omnes}.

\section{Semileptonic $D\to \pi$ and $D\to K$  Decays}
\label{sec:dpik}

As a further test of our predictions for the $B\to \pi$ semileptonic
process, we present results for the $D\to \pi$ and $D\to K$ decays for
which there are precise experimental data~\cite{pdg}:
\begin{eqnarray}
{\cal B}_{\rm exp}(D^0\to \pi^- e^+ \nu_e)&=&(3.6 \pm 0.6)\times
10^{-3},  \qquad \Gamma_{\rm exp} (D^0\to \pi^- e^+ \nu_e) = 
\left (5.8\pm 1.0\right)\times 10^{-12}\mev \label{eq:dpi}\\
{\cal B}_{\rm exp}(D^0\to K^- e^+ \nu_e)&=&(3.58 \pm 0.18)\times
10^{-2},  \quad \Gamma_{\rm exp} (D^0\to K^- e^+ \nu_e) = 
\left (57\pm 3\right)\times 10^{-12}\mev \label{eq:dk},
\end{eqnarray}
with life time $\tau_{D^0}= (410.3\pm 1.5) \times 10^{-15}\,
\mathrm{s}$ and $|V_{cd}|=0.224\pm 0.003$, $|V_{cs}|=0.9737\pm
0.0007$.

In the last two years there has been renewed interest in these decays.
The first three-flavor lattice QCD results~\cite{Latt_D_05} have
appeared, superseding the old quenched ones~\cite{Latt_01,Latt_01_bis,
Latt_D_95}, and the BES~\cite{Exp_D_04} and CLEO~\cite{Exp_D_05}
collaborations have new measurements of the branching ratios
\begin{eqnarray}
{\rm BES:}~{\cal B}(D^0\to \pi^- e^+ \nu_e)&=&(3.3 \pm 1.3)\times
10^{-3}, \quad {\cal B}(D^0\to K^- e^+ \nu_e)=(3.8 \pm 0.5)\times
10^{-2} \\
{\rm CLEO:}~\frac{{\cal B}(D^0\to \pi^- e^+ \nu_e)}{{\cal B}(D^0\to
  K^- e^+ \nu_e)}&=&0.082\pm 0.006 \pm 0.005 \label{eq:cleo}
\end{eqnarray}
Both collaborations have also determined the form
factor at $q^2=0$
\begin{eqnarray}
{\rm BES:}~f^+_\pi(0)&=&0.73\pm 0.14\pm 0.06, \quad f^+_K(0)=0.78\pm
0.04\pm 0.03, \quad \frac{f^+_\pi(0)}{f^+_K(0)}= 0.93 \pm 0.19 \pm
0.07 \label{eq:bes} \\
{\rm CLEO:}~\frac{f^+_\pi(0)}{f^+_K(0)}&=& 0.86 \pm 0.07
^{+0.06}_{-0.04}\pm 0.01 \label{eq:cleo2}
\end{eqnarray}

In the following we will apply the NRCQM developed for the $B\to \pi$
decay to the description of these $D$-meson semileptonic transitions.
All formulae of Sect.~\ref{sec:b2pi} can be used here with the obvious
replacements: $B\to D, B^*\to D^*$ for the $D^0 \to \pi^- l^+ \nu_l$
process, and $B\to D$, $\pi \to K$, $B^* \to D^*_s$ for the $D^0 \to
K^- l^+ \nu_l$ process. We will use $m_{D^0} = 1864.6\mev$,
$m_{D^*}=2010\mev$, $m_{D^*_s}=2112.1\mev$ and $m_{K^-}=493.68\mev$.

\subsection{$D\to\pi l \bar{\nu}_l$}

\begin{figure}
\begin{center}
\makebox[0pt]{\input{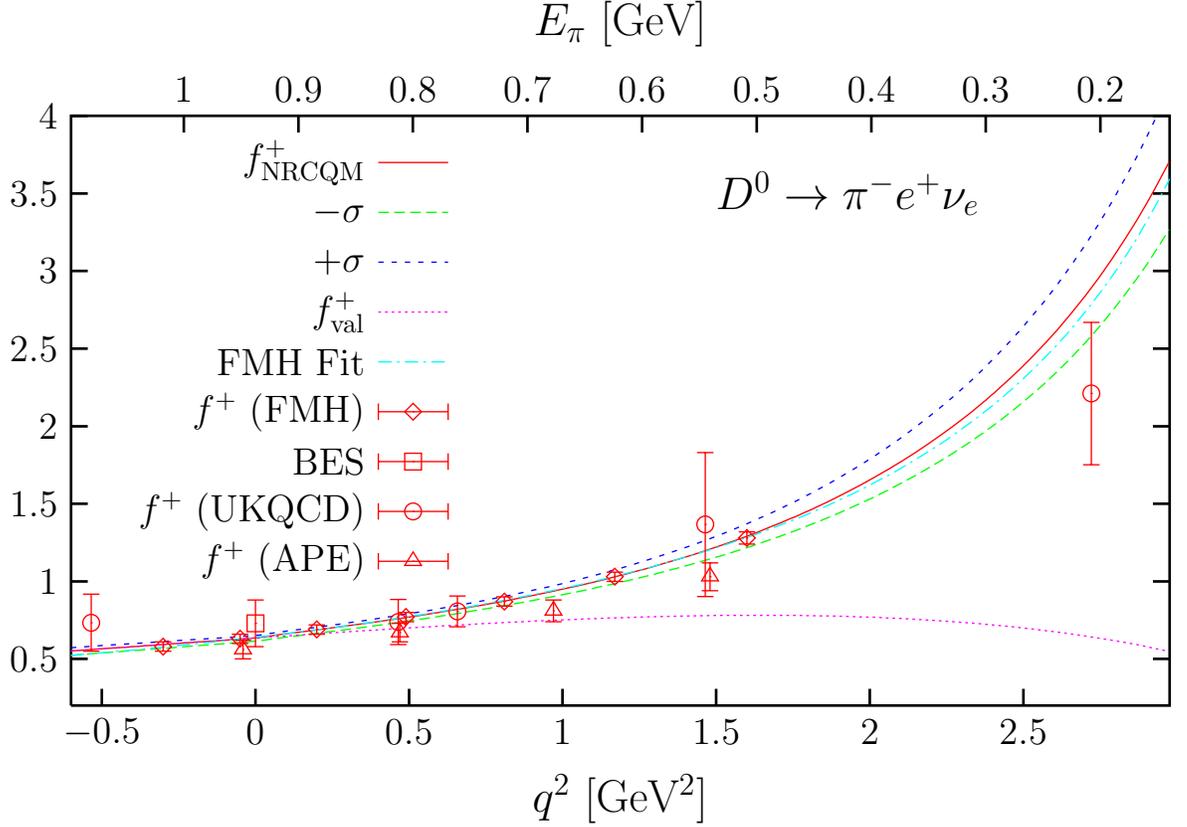}}
\end{center}
\caption{Valence quark (dotted line) and valence quark plus $D^*$-pole
  (solid line denoted $f^+_\mathrm{NRCQM}$) contributions to $f^+$ for
  $D\to \pi$ semileptonic decay. In both cases the AL1 quark-antiquark
  interaction has been used. Triangles (\cite{Latt_01}) and circles
  (\cite{Latt_D_95}) stand for the lattice QCD quenched results
  obtained by the APE and UKQCD Collaborations, respectively. We also
  plot the three-flavor lattice QCD results~\cite{Latt_D_05} from the
  Fermilab-MILC-HPQCD Collaboration (FMH, diamonds), the best fit
  (dash-dotted line) to this latter set of data-points (Eq.~(5) of
  Ref.~\cite{Latt_D_05}), and the determination of $f^+$ at $q^2=0$
  (square) by the BES Collaboration~\cite{Exp_D_04}. Finally, the $\pm
  \sigma$ lines stand for the theoretical uncertainty bands, inherited
  from the errors in Eq.~(\ref{eq:d_err}) and from the quark-antiquark
  interaction model dependence.}
\label{fig:dpi_ff}
\end{figure}

Since there is phase space for the $D^*\to D \pi$ decay to occur, the
$g_{D^*D\pi}$ hadronic constant has been experimentally measured
(CLEO~\cite{gdstar})
\begin{equation}
g_{D^*D\pi}=17.9 \pm 0.3 \pm 1.9 \label{eq:gdstar}
\end{equation}
Taking $f_{D^*}=(234 \pm 20)\mev$ from
Ref.~\protect\cite{Latt_fbstar}, we find\footnote{Note that the
  lattice QCD simulation of Ref.~\cite{gdstar_latt} measured
  $g_{D^*D\pi}=18.8 \pm 2.3 \pm 2.0$ in good agreement with
  Eq.~(\ref{eq:gdstar}).}
\begin{equation}
\left[g_{D^*D\pi}f_{D^*}\right]_{\rm Exp-Latt} = 4.2 \pm 0.6~{\rm GeV}
\label{eq:d_err}
\end{equation}
where we have added errors in quadrature. In Ref.~\cite{NRCQM_def} and
using the same set of NRCQM's, we found a value of $4.9 \pm 0.5 \gev$ for
the above product, in reasonable agreement with
Eq.~(\ref{eq:d_err}). The value quoted in Eq.~(\ref{eq:d_err})
determines the $D^*$-pole contribution, above $q^2=0$, to $f^+$ and
adding it to the valence quark contribution we obtain the results
shown in Fig.~\ref{fig:dpi_ff}. We find excellent agreement between
our description of the form factor and that provided by the unquenched
lattice simulation of Ref.~\protect\cite{Latt_D_05}.  As can be seen
in the figure, the $D^*$-pole contribution is dominant above
$q^2=1.5\gev^2$ and it remains sizeable down to $0.5\gev^2$. We do not
see the need to Omn\`es improve the NRCQM description of the decay
since it is quite good for the whole $q^2$ range. On the other hand,
we see that the pion energy ranges from $m_\pi$ up to about $1\gev$,
which was also the maximum value for $E_\pi$ in the five NRCQM
data-points used in the subtracted Omn\`es representation for the
semileptonic $B\to \pi$ decay depicted in Fig.~\ref{fig:omnes}.  This
reinforces our belief in the reliability of our determination of
$|V_{ub}|$ presented in Subsect.~\ref{sec:err}.
Considering our theoretical uncertainties (errors on
Eq.~(\ref{eq:d_err}) and the spread of results obtained when different
quark-antiquark interactions are considered) together with the
experimental uncertainties on $|V_{cd}|$ quoted above, we find
\begin{equation}
\Gamma_{\rm this~work} (D^0\to \pi^- e^+ \nu_e) = 
\Big [5.2\pm 0.1\,({\rm exp:}|V_{cd}|)\,
 \pm 0.5\,({\rm theory})\Big]\times 10^{-12}\mev
\end{equation}
in good agreement with Eq.~(\ref{eq:dpi}). We also obtain
\begin{equation}
f^+_\pi(0)  = 0.63 \pm 0.02
\end{equation}
compatible within errors with both the BES ($0.73\pm 0.15$) and the
Fermilab-MILC-HPQCD ($0.64 \pm 0.07$) results.

\begin{figure}
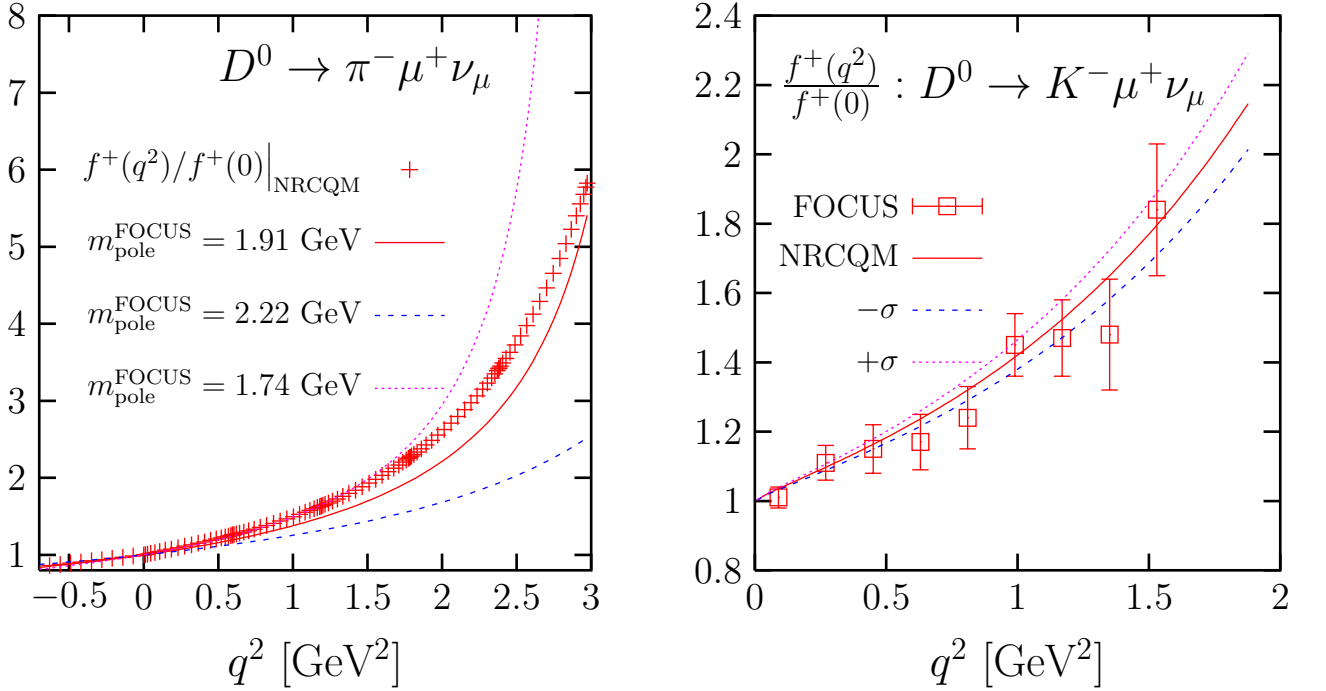

\begin{center}
\makebox[0pt]{\input{focuspi.tex}\hspace{-10mm}\input{focusdk.tex}}
\end{center}
\caption{NRCQM predictions for the ratio $f^+(q^2)/f^+(0)$ for both
$D\to \pi$ (left) and $D\to K$ (right) semileptonic decays.  For
comparison we also plot experimental results from the FOCUS
Collaboration~\cite{focus}: pole fit for the $D\to \pi$ decay
($m_{\rm pole}= 1.91^{+0.31}_{-0.17}\gev$) and direct measurements of the
form factor for different $q^2$ values,  in the $D\to K$ case. In this
latter case, the $\pm \sigma $ lines stand for our theoretical
uncertainty bands, inherited from the errors in
  Eq.~(\ref{eq:d_err}) and from the quark--antiquark
  interaction model dependence.}
\label{fig:focuspi-dk}
\end{figure}

Finally in the left plot of Fig.~\ref{fig:focuspi-dk}, we compare the
NRCQM predictions for the ratio $f^+(q^2)/f^+(0)$ with a pole form
recently fitted to data by the FOCUS Collaboration~\cite{focus}.

\subsection{$D\to K l \bar{\nu}_l$}

Since there is no phase space for the $D^*_s\to D K$ decay, we will
estimate the $g_{D^*_sD K}$ coupling from the value quoted for
$g_{D^*D\pi}$ in Eq.~(\ref{eq:gdstar}). The parameter $\widehat{g}$
defined in Eq.~(\ref{eq:ghat}) describes the strong coupling of
charmed mesons as well as of beauty mesons to the members of the octet
of light pseudoscalars. We will assume flavor SU(3) symmetry for this
basic quantity in the heavy quark chiral effective theory, and thus we
will use~\cite{g_02}
\begin{equation}
g_{D^*_sD K} \approx \frac{2\widehat{g}\sqrt{m_Dm_{D^*_s}}}{f_K} \approx
g_{D^*D \pi} \frac{\sqrt{m_{D^*_s}}}{\sqrt{m_{D^*}}} \frac{f_\pi}{f_K}
\approx 15.3 \pm 1.6 \label{eq:gdk}
\end{equation}
where we have taken $f_K/f_\pi\approx 1.2$ from Ref.~\cite{Latt_fK},
and have kept some SU(3) flavor breaking terms in the masses of the
charmed vector mesons and in the kaon decay constant. Taking
$f_{D^*_s}=(254 \pm 15)\mev$ from Ref.~\cite{Latt_fbstar}, we
find
\begin{equation}
\left[g_{D^*_sDK}f_{D^*_s}\right]_{\rm SU(3)-Latt} = 3.9 \pm 0.5\gev
\label{eq:dk_err}
\end{equation}
\begin{figure}
\begin{center}
\makebox[0pt]{\input{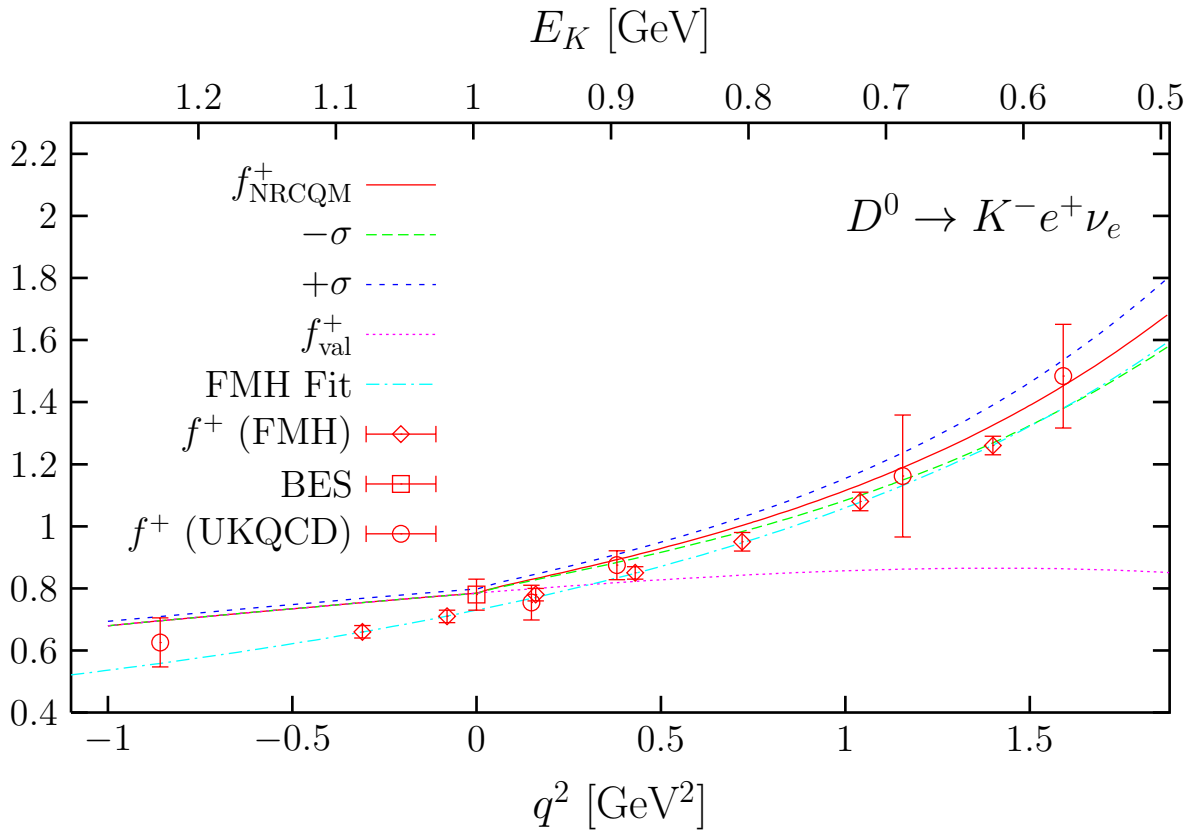}}
\end{center}
\caption{Valence quark (dotted line) and valence quark plus
  $D^*_s$-pole (solid line denoted $f^+_\mathrm{NRCQM}$) contributions
  to $f^+$ for $D\to K$ semileptonic decay. In both cases the AL1
  quark-antiquark interaction has been used. Circles
  (\cite{Latt_D_95}) and diamonds (\cite{Latt_D_05}) stand for the
  lattice QCD quenched and unquenched results obtained by the UKQCD
  and the Fermilab-MILC-HPQCD Collaboration (labelled FMH),
  respectively. We also plot the best fit (dash-dotted line) to this
  latter set of data-points (Eq.~(5) of Ref.~\cite{Latt_D_05}), and
  the determination of $f^+$ at $q^2=0$ (squared) by the BES
  Collaboration~\cite{Exp_D_04}. Finally, the $\pm \sigma$ lines stand
  for the theoretical uncertainty bands, inherited from the errors in
  Eq.~(\ref{eq:d_err}) and from the quark--antiquark interaction model
  dependence.}
\label{fig:dk_ff}
\end{figure}

Our results for this decay are shown in Fig.~\ref{fig:dk_ff}.
Several comments are in order:
\begin{enumerate}
\item We find reasonable agreement with the three-flavor lattice QCD
  results~\cite{Latt_D_05} up to kaon energies of the order of
  $1\gev$, which cover the whole $q^2$ range accessible in the
  physical decay. Discrepancies with lattice data are now more
  sizeable than in the $D\to \pi$ case and lattice unquenched data
  favor values for $\left[g_{D^*_sDK}f_{D^*_s}\right]_{\rm
    SU(3)-Latt}$ smaller than the one used in our calculation (see
  Eq.~(\ref{eq:gdk})). Theoretical errors for $f^+$ are, in this case,
  mostly due to the uncertainties on
  $\left[g_{D^*_sDK}f_{D^*_s}\right]_{\rm SU(3)-Latt}$. Nevertheless,
  we would like to point out that uncertainties on the value of
  $g_{D^*_sDK}$ might be larger that those quoted in
  Eq.~(\ref{eq:gdk}), since flavor SU(3) corrections to the relation
  $g_{D^*_sD K} \approx 2\widehat{g}\sqrt{m_Dm_{D^*_s}}/f_K$ could be
  large ($m_s/m_c \gg m_{d,u}/m_c$).
\item The contribution of the vector resonance is  less
  important than in the $B \to \pi$ and $D \to \pi$ decays, since the
  $D^*_s$  is located relatively far from $\sqrt{q^2_{\rm max}}$. 
\item Our predictions for $f^+$ at negative values of $q^2$, which do
  not enter into the phase space integral, suffer from larger
  uncertainties, since in that region the transferred momentum is
  larger than $1\gev$ and, as for the $B\to \pi$ case, relativistic
  effects could became important. One could Omn\`es improve the NRCQM
  to achieve a better description of the form factor in the negative
  $q^2$ region.
\end{enumerate}
In the right plot of Fig.~\ref{fig:focuspi-dk}, we compare the NRCQM
predictions for the ratio $f^+(q^2)/f^+(0)$ with recently measured
data from the FOCUS Collaboration~\cite{focus} and find satisfactory
and reassuring agreement.

For the integrated width, we find
\begin{equation}
\Gamma_{\rm this~work} (D^0\to K^- e^+ \nu_e) = 
\Big [66\pm  3\,({\rm theory})\Big]\times 10^{-12}~{\rm MeV}
\end{equation}
which is about two standard deviations higher than the value quoted in
Eq.~(\ref{eq:dk}).

Eqs.~(\ref{eq:dpi}) and~(\ref{eq:dk}) lead to (adding errors in
quadratures)
\begin{equation}
\frac{{\cal B}(D^0\to \pi^- e^+ \nu_e)}{{\cal B}(D^0\to
  K^- e^+ \nu_e)}= 0.101 \pm 0.017
\end{equation}
which turns out to be a bit higher, though compatible within errors,
than the recent CLEO determination quoted in Eq.~(\ref{eq:cleo}). For
this ratio of branching fractions we find
\begin{equation}
\frac{{\cal B}(D^0\to \pi^- e^+ \nu_e)}{{\cal B}(D^0\to
  K^- e^+ \nu_e)}\Big|_{\rm this~work}= 0.079 \pm 0.008
\end{equation}
in excellent agreement with the CLEO measurement. We also find
\begin{equation}
f^+_K(0)  = 0.79 \pm 0.01, \quad \frac{f^+_\pi(0)}{f^+_K(0)} = 0.80
\pm 0.03
\end{equation}
which compare well to the recent experimental measurements in
Eqs.~(\ref{eq:bes}) and~(\ref{eq:cleo2}).

\section{Concluding Remarks}
\label{sec:concl}

We have shown the limitations of a valence quark model to describe the
$B\to \pi$, $D\to \pi$ and $D\to K$ semileptonic decays. As a first
correction, we have included in each case the heavy--light vector
resonance pole contribution. For the semileptonic $B\to \pi$ decay,
the inclusion of the $B^*$ degree of freedom provides a realistic
$q^2$ dependence of the relevant form factor, $f^+$, from $q^2_{\rm
  max}$ down to around $18\gev^2$. We then use a multiply-subtracted
Omn\`es dispersion relation, which considerably diminishes the form
factor dependence on the elastic $\pi B \to \pi B$ scattering
amplitudes at high energies, to combine LCSR results at $q^2=0$ with
NRCQM predictions in the high $q^2$ region. As a result we have been
able to predict the $f^+$ form factor for all $q^2$ values accessible
in the physical decay. We have used a Monte Carlo procedure and
analyzed the predictions of five different quark-antiquark
interactions to determine theoretical error bands for form factors and
the decay width. This has allowed us to extract from the measured
branching fraction the value $|V_{ub}|= 0.0034 \pm 0.0003\, ({\rm
  exp}) \pm 0.0007\, ({\rm theory})$ in excellent agreement with the
CLEO Collaboration determination of Ref.~\cite{Exp_03}. For the $D\to
\pi$ semileptonic decay we have found excellent agreement between our
model calculation (valence quark plus $D^*$-pole contributions) of
$f^+$ and the one obtained by the unquenched lattice simulation of
Ref.~\cite{Latt_D_05}. We found no need to Omn\`es-improve our
calculation in this case. Our results $\Gamma(D^0\to \pi^-e^+\nu_e)=
\left[5.2\pm 0.1\,({\rm exp:}|V_{cd}|)\, \pm 0.5\,({\rm
    theory})\right] \times 10^{-12}~{\rm MeV}$ and $f^+_\pi(0) = 0.63
\pm 0.02$ are in good agreement with experimental data. Finally for
the $D\to K$ semileptonic decay we find good agreement, in the
physical region, between our model calculation (valence quark plus
$D^*_s$-pole contributions) of $f^+$ and lattice data, from
UKQCD~\cite{Latt_D_95} and Fermilab-MILC-HPQCD~\cite{Latt_D_05}, and
also with recent measurements from the FOCUS
Collaboration~\cite{focus}. Again our results ${{\cal B}(D^0\to \pi^-
  e^+ \nu_e)}/{{\cal B}(D^0\to K^- e^+ \nu_e)}= 0.079 \pm 0.008 $,
$f^+_K(0) = 0.79 \pm 0.01$ and ${f^+_\pi(0)}/{f^+_K(0)} = 0.80 \pm
0.03$ are in good agreement with experimental determinations by the
CLEO~\cite{Exp_D_05} and BES~\cite{Exp_D_04} Collaborations.

\begin{acknowledgments}
This research was supported by DGI and FEDER funds, under contracts
BFM2002-03218, BFM2003-00856 and FPA2004-05616, by the Junta de
Andaluc\'\i a and Junta de Castilla y Le\'on under contracts FQM0225
and SA104/04, and it is part of the EU integrated infrastructure
initiative Hadron Physics Project under contract number
RII3-CT-2004-506078. C. Albertus wishes to acknowledge a grant from
Junta de Andaluc\'\i a. J.M. Verde-Velasco acknowledges a grant
(AP2003-4147) from the Spanish Ministerio de Educaci\'on y Ciencia.
J.M.~Flynn acknowledges PPARC grant PPA/G/O/2002/00468, the
hospitality of the Universidad de Granada and the Institute for
Nuclear Theory at the University of Washington, and thanks the
Department of Energy for partial support during the completion of this
work.
\end{acknowledgments}

\appendix 

\section{Multiply Subtracted Omn\`es Dispersion Relation}

Let the form factor\footnote{The discussion below can be trivially
  generalized to any scattering amplitude or form factor with definite
  total angular momentum and isospin quantum numbers.} $f^+(s)$ be
analytic on the complex $s$ plane (Mandelstam's hypothesis~\cite{Ma58}
of maximum analyticity) except for a cut $L\equiv
[\sth=(m_B+m_\pi)^2,+\infty[$ along the real positive $s$ axis, as
    demanded by Watson's theorem~\cite{Wa55}. For real values $s<
    \sth$ the form factor is real which implies that the values of
    the form factor above and below the cut are complex conjugates of
    each other: $ f^+(s+\mi\epsilon) = f^+(s-\mi\epsilon)^*$. For $s\ge
    s_{\rm th}$, the form factor has a discontinuity across the cut
    and develops an imaginary part $f^+(s+\mi\epsilon) -
    f^+(s-\mi\epsilon) = 2\mi{\rm Im} f(s+\mi\epsilon)$. Cauchy's theorem
    implies that $f^+(s)$ can be written as a dispersive integral
    along the cut and performing one subtraction at $s_0 <s_{\rm th}$
    one gets:
\begin{equation}
f^+(s) = f^+(s_0)+\frac{s-s_0}{\pi} \int_{\sth}^{+\infty}
    \frac{dx}{x-s_0} \frac{{\rm Im}f^+(x)}{x-s},
    \quad s \notin L, \quad  s_0 < s_{\rm th} \label{eq:omn}
\end{equation} 
Depending on the asymptotic behavior of $f^+(s)$ at the extremes of
the cut $L$, more subtractions may be needed to make the integral
convergent. For the time being, let us assume that one subtraction is
sufficient. The well known Omn\`es solution for the above dispersive
representation is~\cite{Omnes}:
\begin{equation}
O(s) = f^+(s_0) \exp\left \{\frac{s-s_0}{\pi} \int_{\sth}^{+\infty}
    \frac{dx}{x-s_0} \frac{\delta(x)}{x-s}  \right \},
    \quad s \notin L, \quad  s_0 < s_{\rm th} \label{eq:omn2}
\end{equation} 
with $\delta(s)$ the elastic $\pi B \to \pi B$ phase
shift\footnote{Obviously $\delta(s)$ has to be defined as a continuous
function of $s$.} in the
$J^P=1^-$ and isospin $1/2$ channel (see Eq.~(\ref{eq:wat})). $O(s)$
gives the physical form-factor since,
\begin{enumerate}
\item For $s\ge \sth$, we have
\begin{eqnarray}
O(s\pm \mi\epsilon) &=&  f^+(s_0)\exp\left
    \{\frac{s-s_0}{\pi}\left [\, {\cal P}\int_{\sth}^{+\infty}
    \frac{dx}{x-s_0} \frac{\delta(x)}{x-s} \pm
 \mi \pi \frac{\delta(s)}{s-s_0}\right ]\right\} \nonumber \\
&=& e^{\pm \mi\delta(s)} \left [ f^+(s_0)\exp\left
    \{\frac{s-s_0}{\pi}\, {\cal P}\int_{\sth}^{+\infty}
    \frac{dx}{x-s_0}   \frac{\delta(x)}{x-s}   \right\} \right ]
\end{eqnarray}
where ${\cal P}$ stands for principal part of the integral. Thus we
have $ O(s+\mi\epsilon) = O(s-\mi\epsilon)^*$, the function $O$ is real
for $s<\sth$ and has neither poles nor cuts, except for that
required by Watson's theorem: $L\equiv
[\sth,+\infty[$.  The discontinuity across this cut is given by
$O(s+\mi\epsilon) - O(s-\mi\epsilon) =2\mi{\rm Im}
O(s+\mi\epsilon)$ and by construction $O(s_0)=f^+(s_0)$.  Thus,
both $f^+(s)$ and $O(s)$ satisfy the same dispersion relation
(Eq.~(\ref{eq:omn})) and therefore both functions can differ at
most by a polynomial with real coefficients, which should vanish at
$s=s_0$. But, this polynomial is zero since:

\item The function $O(s)$ satisfies Watson's theorem:
\begin{equation}
{ O(s+ \mi  \epsilon ) \over O(s- \mi  \epsilon ) }
 = e^{2\mi\delta(s)}
 = { f^+(s+ \mi  \epsilon ) \over f^+(s- \mi  \epsilon ) } , \quad s> \sth
\end{equation}
\end{enumerate}
Performing $n+1$ subtractions, one can can produce a rank $n+2$
polynomial in the denominator of the dispersive integral of
Eq.~(\ref{eq:omn}). Indeed, for $ s \notin L$
\begin{equation}
f^+(s)= P_n(s)+\frac{(s-s_0)(s-s_1)\cdots(s-s_n)}{\pi}
    \int_{\sth}^{+\infty} \frac{dx}{(x-s_0)(x-s_1)\cdots(x-s_n)}
    \frac{{\rm Im}f^+(x)}{x-s}, \quad s_0,
    \cdots,s_n < s_{\rm th} \label{eq:multiply}
\end{equation}
with the rank $n$ polynomial $P_n(s)$ determined by the $n+1$
equations $ P_n(s_i) = f^+(s_i),~i=0,1,\cdots,n$,
\begin{equation}
P_n(s) = \sum_{j=0}^n \alpha_j(s) f^+(s_j), \qquad \alpha_j(s) =
\left[ \prod_{j\ne k=0}^{n} \frac{s-s_k}{s_j-s_k} \right]
\end{equation}
Note that $\alpha_j(s)$ are rank $n$ polynomials, which satisfy
\begin{equation}
\sum_{j=0}^n \alpha_j(s)  = 1 \label{eq:alphas}
\end{equation}
On the other hand for $s>s_{\rm th}$, we have from Eq.~(\ref{eq:wat})
\begin{equation}
\log f^+(s+{\mi \epsilon}) - \log f^+(s-{\mi \epsilon}) = \log  {
  f^+(s+ \mi  \epsilon ) \over f^+(s- \mi  \epsilon ) } =
  2 \mi \delta(s)  = 2 \mi {\rm Im} \left [ \log f^+(s+{\mi
  \epsilon})\right]
\end{equation}
Thus in analogy to Eq.~(\ref{eq:multiply}), assuming that the form
factor does not vanish in ${\cal C}-\{\sth\}$\footnote{Note that, we
have already treated $s=\sth$ as a branch point.}, or neglecting the
contribution from the log cut if it has a finite branch point
different from $\sth$, we can write
\begin{equation}
\log f^+(s)= \widehat{P}_n(s)+\frac{(s-s_0)(s-s_1)\cdots(s-s_n)}{\pi}
    \int_{\sth}^{+\infty} \frac{dx}{(x-s_0)(x-s_1)\cdots(x-s_n)}
    \frac{\delta(x)}{x-s}, \quad  s \notin L
\end{equation}
with
%
\begin{equation}
\widehat{P}_n(s) = \sum_{j=0}^n \alpha_j(s) \log f^+(s_j)
\end{equation}
From the above equation one readily finds the $(n+1)$-subtracted
Omn\`es representation given in Eq.~(\ref{eq:omnes}).

Finally, we would like to draw the reader's attention to a subtle
point. In Eq.~(\ref{eq:omn}) we have assumed that $f^+$ has no
poles. However we know that if the scattering amplitude has a pole at
$s_R=M_R^2-\mi M_R\Gamma_R$ on its second Riemann sheet (resonance) or on
the physical sheet (bound state with $\Gamma_R=0^+$ and $M_R^2 <
\sth$.), it might show up as a pole in the complex plane of $f^+$ (see
Eq.~(\ref{eq:polefg})). On the other hand, the $S$-matrix depends on
$\exp{(2\mi\delta)}$, and thus one has the freedom to add factors of
$m\pi$, for $m$ an integer, to the phase-shift without modifying the
$S$-matrix. However, the Omn\`es representation of the form factor
will definitely depend on the specific value chosen for the integer
$m$.

To fix this ambiguity, we will assume that at threshold the phase
shift should be $\delta(\sth) = n_b \pi$, where $n_b$ is the number of
bound states in the channel, while $\delta (\infty) = k \pi$, where $k$
is the number of zeros of the scattering amplitude on the physical
sheet (this is Levinson's theorem~\cite{MS70}). This choice for the
phase shifts also takes into account the existence of poles in the
scattering matrix. We demonstrate with a simple example in which a
$p$-wave $T$-matrix is proportional to
$(s-\sth)/(s-M_R^2+\mi M_R\Gamma_R)$. The phase shift is given by
\begin{equation}
\delta(s) = \pi + {\rm
  Arctan}\left[\frac{-M_R\Gamma_R}{s-M_R^2}\right], \qquad s > \sth
\end{equation}
with ${\rm Arctan} \in [-\pi,\pi[$. This satisfies $\delta (\infty) =
    \pi$ and, if $M_R \Gamma_R \ll |s-M_R^2|$, it also leads to
    $\delta(\sth) = \pi$ or $0$ for a bound state or resonance
    respectively, in accordance with Levinson's theorem. For
    simplicity, let us also assume $\Gamma_R \ll M_R$. In this
    circumstance we can approximate
\begin{equation}
\delta(s) \approx \pi \left[1 - H(M^2_R-s)\right] = \pi H(s-M^2_R)
\end{equation}
where $H(\;)$ is the step function. Since
\begin{equation}
\frac{s-s_0}{\pi} \int_{\sth}^{+\infty} \frac{dx}{x-s_0}
    \frac{\delta(x)}{x-s} \approx (s-s_0)
    \int_{{\rm Max}(\sth,M_R^2)}^{+\infty} \frac{dx}{x-s_0}
    \frac{1}{x-s} = \log\left\{\frac{{\rm Max}(\sth,M_R^2)-s_0}{{\rm
    Max}(\sth,M_R^2)-s}\right\},
\end{equation}
we find that the Omn\`es solution from a once-subtracted dispersion
relation, Eq.~(\ref{eq:omn2}), reads
\begin{equation}
O(s) \approx f^+(s_0) \frac{{\rm Max}(\sth,M_R^2)-s_0}{{\rm
    Max}(\sth,M_R^2)-s}
\end{equation}
It has a pole at $s=M^2_R$ for a resonance or at $s=\sth$ in the case
of a bound state. For the case of a resonance with a finite width, the
pole will move to $s=M^2_R + \mi M_R \Gamma_R$. On the other hand, for a
bound state, going beyond the approximation $\delta(s)\approx \pi$
(see Eq.~(\ref{eq:levin})), the form factor will be sensitive to the
exact position of the pole ($s=M_R^2$), since the effective range
parameters (scattering volume, \ldots) will depend on $M_R$.

These conclusions can easily be generalized when a multiply-subtracted
Omn\`es dispersion relation is used.

\end{document}